\documentclass[11pt]{revtex4} 
\usepackage{amssymb,epsf}
\usepackage{latexsym}

\begin{document}

\title{Magnetic Branes in Brans-Dicke-Maxwell Theory }
\author{A. Sheykhi$^{1,2}$\footnote{sheykhi@mail.uk.ac.ir} and
E. Ebrahimi$^{1,3}$}
\address{$^1$Department of Physics, Shahid Bahonar University, P.O. Box 76175, Kerman,  Iran\\
         $^2$Research Institute for Astronomy and Astrophysics of Maragha (RIAAM), Maragha, Iran\\
         $^3$Physics Department and Biruni Observatory, Shiraz University, Shiraz 71454, Iran}
\vspace*{1.5cm}
\begin{abstract}
\vspace*{1.5cm}\centerline{\bf Abstract} We present a new class of
magnetic brane solutions in $(n+1)$-dimensional
Brans-Dicke-Maxwell theory in the presence of a quadratic
potential for the scalar field. These solutions are neither
asymptotically flat nor (anti)-de Sitter. Our strategy for
constructing these solutions is applying a conformal
transformation to the corresponding solutions in dilaton gravity.
This class of solutions represents a spacetime with a longitudinal
magnetic field generated by a static brane. They have no curvature
singularity and no horizons but have a conic geometry with a
deficit angle $\delta$. We generalize this class of solutions to
the case of spinning magnetic brane with all rotation parameters.
We also use the counterterm method and calculate the conserved
quantities of the solutions.
\end{abstract}
\pacs{04.20.Jb, 04.40.Nr, 04.50.+h}
 \maketitle
 \section{Introduction\label{Intr}}

It seems likely  that the standard model of cosmology, based on
Einstein gravity, could not describe the acceleration of the
universe expansion correctly \cite{Rie}. Thus, cosmologists have
attended to alternative theories of gravity to explain the
accelerated expansion. One of the alternative theories of general
relativity, that arose a lot of enthusiasm recently, is the
scalar-tensor theory. Scalar-tensor theories are not new and have
a long history. The pioneering study on scalar-tensor theories was
done by Brans and Dicke several decades ago who sought to
incorporate Mach's principle into gravity \cite{BD}. According to
Brans-Dicke (BD) theory the phenomenon of inertia arises from
acceleration with respect to the general mass distribution of the
universe. This theory can be regarded as an economic modification
of general relativity which accomodates both Mach's principle and
Dirac's large number hypothesis as new ingredients. This theory is
self-consistent, complete and for $\omega>10^4$  is consistent
with solar system observations, where $\omega$ is a coupling
parameter \cite{Will,Bert,Aca}. In recent years this theory got a
new impetus as it arises naturally as the low energy limit of many
theories of quantum gravity such as superstring theory or
Kaluza-Klein theory. In string theory, gravity becomes
scalar-tensor in nature. The low-energy effective action of the
superstring theory leads to the Einstein gravity, coupled
non-minimally to a scalar field. Due to highly nonlinear character
of BD theory, a desirable pre-requisite for studying strong field
situation is to have knowledge of exact explicit solutions of the
field equations. And as black holes are very important both in
classical and quantum gravity, many authors have investigated
various aspects of them in BD theory \cite{Sen}. It turned out
that the dynamic scalar field in the BD theory plays an important
role in the process of collapse and critical phenomenon. The study
on the black hole solutions in BD theory have been carried out
extensively in  the literature \cite{Brans, Hawking,
Cai1,Cai2,Kim,Deh1,Gao,SheyAla}.

Besides investigating various aspects of black hole solutions in
BD theory, there has been a lot of interest in recent years, in
studying the horizonless solutions in various gravity theories.
Strong motivation for studying such kinds of solutions comes from
the fact that they may be interpreted as cosmic strings. Cosmic
strings are topological defects which are inevitably formed during
phase transitions in the early universe, and their subsequent
evolution and observational signatures must therefore be
understood. The string model of structure formation may help to
resolve one of cosmological mystery, the origin of cosmic magnetic
fields \cite{TV}. There is strong evidence from all numerical
simulations for the scaling behavior of the long string network
during the radiation-dominated era. Apart from their possible
astrophysical roles, topological defects are fascinating objects
in their own right. Their properties, which are very different
from those of more familiar system, can give rise to a rich
variety of unusual mathematical and physical phenomena \cite{AV}.
A short review of papers treating this subject follows. The
four-dimensional horizonless solutions of Einstein gravity have
been explored in \cite{Vil,Ban}. These horizonless solutions
\cite{Vil,Ban} have a conical geometry; they are everywhere flat
except at the location of the line source. The spacetime can be
obtained from the flat spacetime by cutting out a wedge and
identifying its edges. The wedge has an opening angle which turns
to be proportional to the source mass. The extension to include
the Maxwell field \cite{Bon} and the cosmological constant
\cite{Lem1,Lem2} have also been done. The generalization of these
asymptotically AdS magnetic rotating solutions of the
Einstein-Maxwell equation to higher dimensions \cite{Deh2} and
higher derivative gravity \cite{Deh3} have been also done. In the
context of electromagnetic cosmic string, it has been shown that
there are cosmic strings, known as superconducting cosmic strings,
that behave as superconductors and have interesting interactions
with astrophysical magnetic fields \cite{Wit2}. The properties of
these superconducting cosmic strings have been investigated in
\cite{Moss}. It is also of great interest to generalize the study
to the scalar-tensor theory. Attempts to explore the physical
properties of the horizonless solutions in dilaton gravity
\cite{Fer,Deh4,SDR,DSH,Sheykhi} and three \cite{Dia} and four
\cite{Sen1} dimensional BD theory have also been done. Our aim in
this paper is to construct $(n+1)$-dimensional horizonless
solutions of Brans-Dicke-Maxwell (BDM) theory for an arbitrary
value of coupling constant and investigate the effects of the
scalar field on the properties of the spacetime such as the
deficit angle of the spacetime.

The structure of our paper is as follows. In section \ref{Conf},
we present the basic equations and the conformal transformation
between the action of the dilaton gravity theory and the BD
theory. In section \ref{Static}, we present the metric for the
$(n+1)$-dimensional magnetic solutions in BDM theory with
longitudinal magnetic field generated by a static source. In
section \ref{Spin}, we generalize these solutions to the case of
spining branes. In section \ref{Cons}, we obtain the conserved
quantities of the spacetimes through the use of the counterterm
method. The last section is devoted to conclusion.

\section{ Field equations and Conformal Transformations}\label{Conf}
The action of the $(n+1)$-dimensional BDM theory with one scalar
field $\Phi$ and a self-interacting potential $V(\Phi)$ can be
written as
\begin{equation}
I_{G}=-\frac{1}{16\pi}\int_{\mathcal{M}}
d^{n+1}x\sqrt{-g}\left(\Phi {R}-\frac{\omega}{\Phi}(\nabla\Phi)^2
-V(\Phi)-F_{\mu \nu}F^{\mu \nu}\right),\label{act1}
\end{equation}
where ${R}$ is the scalar curvature, $V(\Phi )$ is a potential for
the scalar field $\Phi $, $F_{\mu \nu }=\partial _{\mu }A_{\nu
}-\partial _{\nu }A_{\mu }$ is the electromagnetic field tensor,
and $A_{\mu }$ is the electromagnetic potential. The factor
$\omega$ is the coupling constant. We obtain the equations of
motion by varying the action (\ref{act1}) with respect to the
gravitational field $g_{\mu \nu }$, the scalar field $\Phi $ and
the gauge field $A_{\mu }$. The result is
\begin{eqnarray}
&&G_{\mu
\nu}=\frac{\omega}{\Phi^2}\left(\nabla_{\mu}\Phi\nabla_{\nu}\Phi-\frac{1}{2}g_{\mu
\nu}(\nabla\Phi)^2\right)
-\frac{V(\Phi)}{2\Phi}g_{\mu \nu}+\frac{1}{\Phi}\left(\nabla_{\mu}\nabla_{\nu}\Phi-g_{\mu \nu}\nabla^2\Phi\right)\nonumber \\
&&+\frac{2}{\Phi}\left(F_{\mu \lambda}F_{ \nu}^{\
\lambda}-\frac{1}{4}F_{\rho \sigma}F^{\rho
\sigma}g_{\mu \nu}\right), \label{Eq1}\\
&&\nabla^2\Phi=-\frac{n-3}{2(n-1)\omega+2n}F^2+\frac{1}{2(n-1)\omega+2n}\left((n-1)\Phi\frac{dV(\Phi)}{d\Phi}
-(n+1)V(\Phi)\right),\label{Eq2} \\
&&\nabla_{\mu}F^{\mu \nu}=0, \label{Eq3}
\end{eqnarray}
where $G_{\mu \nu}$ and $\nabla$ are, respectively, the Einstein
tensor and covariant differentiation in the spacetime metric
$g_{\mu \nu}$. It is clear that the right hand side of Eq.
(\ref{Eq1}) includes the second derivatives of the scalar field,
so it is hard to solve the field equations (\ref{Eq1})-(\ref{Eq3})
directly. Fortunately, we can remove this difficulty by a
conformal transformation. Indeed, the BDM theory (\ref{act1}) can
be transformed into the Einstein-Maxwell-dilaton theory via the
conformal transformation
\begin{eqnarray} \label{conf}
&&\bar{g}_{\mu \nu}=\Omega^ {-2}g_{\mu \nu},
\end{eqnarray}
with
\begin{eqnarray}
&&\Omega^ {-2}=\Phi^{\frac{2}{n-1}},
\end{eqnarray}
and
\begin{equation}
\alpha=\frac{n-3}{\sqrt{4(n-1)\omega+4n}}, \  \   \
\bar{\Phi}=\frac{n-3}{4\alpha}\ln \Phi. \label{6}
\end{equation}
Using this conformal transformation, the action (\ref{act1})
transforms to
\begin{equation}
\bar{I}_{G}=-\frac{1}{16\pi}\int_{\mathcal{M}}
d^{n+1}x\sqrt{-\bar{g}}\left({\bar{R}}-\frac{4}{n-1}(\bar{\nabla}\
\bar{\Phi})^2-\bar{V}(\bar{\Phi})-e^{-\frac{4\alpha\bar{\Phi}}{n-1}}\bar{F}_{\mu
\nu}\bar{F}^{\mu \nu}\right), \label{act2}
\end{equation}
where ${\bar{R}}$ and $\bar{\nabla}$ are, respectively, the Ricci
scalar and covariant differentiation in the spacetime metric
$\bar{g}_{\mu \nu}$, and $\bar{V}(\bar{\Phi})$ is
\begin{equation}
\bar{V}(\bar{\Phi})=\Phi^{-\frac{n+1}{n-1}}V(\Phi).\label{8}
\end{equation}
This action is just the action of the $(n+1)$-dimensional
Einstein-Maxwell-dilaton gravity, where $\bar{\Phi}$ is the
dilaton field and $\bar{V}(\bar{\Phi})$ is a potential for
$\bar{\Phi}$. $\alpha $ is an arbitrary constant governing the
strength of the coupling between the dilaton and the Maxwell
field. Varying the action (\ref{act2}), we can obtain equations of
motion
\begin{eqnarray}
&&\bar{{R}}_{\mu
\nu}=\frac{4}{n-1}\left(\bar{\nabla}_{\mu}\bar{\Phi}\bar{\nabla}
_{\nu}\bar{\Phi}+\frac{1}{4}\bar{V}(\bar{\Phi})\bar{g}_{\mu
\nu}\right)+ 2e^{\frac{-4\alpha\bar{\Phi}}{n-1}}\left(\bar{F}_{\mu
\lambda}\bar{F}_{\nu}^{ \ \lambda} -\frac{1}{2(n-1)}\bar{F}_{\rho
\sigma}\bar{F}^{\rho \sigma}\bar{g}_{\mu \nu}\right), \label{Eqd1}\\
&&
\bar{\nabla}^2\bar{\Phi}=\frac{n-1}{8}\frac{\partial\bar{V}}{\partial\bar{\Phi}}
-\frac{\alpha}{2}e^{\frac{-4\alpha\bar{\Phi}}{n-1}}\bar{F}_{\rho
\sigma}\bar{F}^{\rho \sigma},\label{Eqd2}\\
&&
\bar{\nabla}_{\mu}\left(e^{\frac{-4\alpha\bar{\Phi}}{n-1}}\bar{F}^{\mu
\nu}\right)=0. \label{Eqd3}
\end{eqnarray}
Comparing Eqs. (\ref{Eq1})-(\ref{Eq3}) with Eqs.
(\ref{Eqd1})-(\ref{Eqd3}), we find that if $\left(\bar{g}_{\mu
\nu},\bar{F}_{\mu \nu},\bar{\Phi}\right)$ is the solution of Eqs.
(\ref{Eqd1})-(\ref{Eqd3}) with potential $\bar{V}(\bar{\Phi})$,
then
\begin{equation}\label{conform2}
\left[{g}_{\mu \nu},{F}_{\mu
\nu},{\Phi}\right]=\left[\exp\left({\frac{-8\alpha
\bar{\Phi}}{(n-1)(n-3)}}\right)\bar{g}_{\mu \nu},\bar{F}_{\mu
\nu},\exp\left({\frac{4\alpha \bar{\Phi}}{n-3}}\right)\right],
\end{equation}
is the solution of Eqs. (\ref{Eq1})-(\ref{Eq3}) with potential
$V(\Phi)$. In this paper we consider the action (\ref{act1}) with
a quadratic potential
\[
V(\Phi)=2\Lambda\Phi^2.
\]
Applying the conformal transformation (\ref{conform2}), we obtain
the potential $ \bar{V}(\bar\Phi)$ in the dilaton gravity theory
\begin{equation}
\bar{V}(\bar\Phi)=2\Lambda
e^{\frac{4\alpha\bar{\Phi}}{n-1}},\label{a1}
\end{equation}
which is a Liouville-type potential. Therefore, instead of solving
the complicated Eqs. (\ref{Eq1})-(\ref{Eq3}) of BD theory with
quadratic potential, we solve the corresponding Eqs.
(\ref{Eqd1})-(\ref{Eqd3}) of the dilaton gravity theory with a
Liouville-type potential. Then, by applying  the conformal
transformations (\ref{conform2}) we obtain magnetic brane
solutions in BD theory.
\section{Static Magnetic Branes in BD theory\label{Static}}
Here we want to obtain the $(n+1)$-dimensional solutions of Eqs.
(\ref{Eq1})-(\ref{Eq3}) which produce longitudinal magnetic fields
in the Euclidean submanifold spans by $x^{i}$ coordinates
($i=1,...,n-2$). We assume the following form for the metric
\begin{eqnarray}
d\bar{s}^{2}=-\frac{\rho ^{2}}{l^{2}}R^{2}(\rho )dt^{2}+\frac{d\rho ^{2}}{f(\rho )}%
+l^{2}f(\rho )d\phi ^{2}+\frac{\rho ^{2}}{l^{2}}R^{2}(\rho
)dX^{2}{,}
 \label{met1}
\end{eqnarray}
where $dX^{2}={{\sum_{i=1}^{n-2}}}(dx^{i})^{2}$ is the Euclidean
metric on the
$(n-2)$-dimensional submanifold. The coordinates $%
x^{i}$'s have dimension of length and range in $(-\infty,\infty)$,
while the angular coordinate $\phi $ is dimensionless as usual and
ranges in $[0,2\pi] $. The motivation for this metric gauge
$[g_{tt}\varpropto -\rho ^{2}$ and $(g_{\rho \rho
})^{-1}\varpropto g_{\phi \phi }]$ instead of the usual Schwarzschild gauge $%
[(g_{\rho \rho })^{-1}\varpropto g_{tt}$ and $g_{\phi \phi
}\varpropto \rho ^{2}]$ comes from the fact that we are looking
for a horizonless solution \cite{Lem2}. Inserting metric
(\ref{met1}) in the field equations (\ref{Eqd1})-(\ref{Eqd3}) of
the dilaton gravity theory, one can show that these equations have
solutions of the form \cite{SDR}
\begin{eqnarray}
&&f(\rho )=\frac{2\Lambda (\alpha ^{2}+1)^{2}b^{2\gamma }}{(n-1)(\alpha ^{2}-n)%
}\rho ^{2(1-\gamma )}+\frac{m}{\rho ^{(n-1)(1-\gamma )-1}}-\frac{%
2q^{2}(\alpha ^{2}+1)^{2}b^{-2(n-2)\gamma }}{(n-1)(\alpha
^{2}+n-2)\rho ^{2(n-2)(1-\gamma )}},\nonumber \\ && \bar{\Phi}
(\rho )=\frac{(n-1)\alpha }{2(1+\alpha ^{2})}\ln (\frac{b}{\rho
}), \label{Phir} \nonumber \\ && \label{Rphi} R(\rho
)=e^{2\alpha\bar{\Phi} /(n-1)}, \nonumber \\ && \label{Ftr}
\bar{F}_{\phi \rho}=\frac{ql e^{4\alpha \bar{\Phi} /(n-1)}}{(\rho
R)^{n-1}}.
\end{eqnarray}
where $\gamma=\alpha^2/(\alpha^2+1)$, $q$ is the charge parameter
of the brane, and  $b$ and $m$ are arbitrary constants. Applying
the conformal transformation (\ref{conform2}), the
$(n+1)$-dimensional magnetic solutions of BD theory can be
obtained as
\begin{eqnarray}
ds^{2}=-\frac{\rho ^{2}}{l^{2}}H^{2}(\rho )dt^{2}+\frac{d\rho ^{2}}{V(\rho )}%
+l^{2}U(\rho )d\phi ^{2}+\frac{\rho ^{2}}{l^{2}}H^{2}(\rho
)dX^{2}{,}
 \label{met2}
\end{eqnarray}
where $U(\rho)$, $V(\rho)$, $H(\rho)$ and $\Phi(\rho)$ are
\begin{eqnarray}
U(\rho)&=&\frac{2\Lambda(\alpha^2+1)^{2}b^{2\gamma(\frac{n-5}{n-3})}}{(n-1)(\alpha^2-n)}\rho^{2(1-\frac{\gamma(n-5)}{n-3})}+
\frac{mb^{(\frac{-4\gamma}{n-3})}}{\rho^{n-2}}\rho^{\gamma\left(n-1+\frac{4}{n-3}\right)}\nonumber \\
&&-\frac{2q^{2}(\alpha^2+1)^{2}b^{-2\gamma\left(n-2+\frac{2}{n-3}\right)}}{(n-1)\left(\alpha^2+n-2\right)\rho^{2[(n-2)(1-\gamma)-
\frac{2\gamma}{n-3}]}},
 \label{U1}\\
V(\rho)&=&\frac{2\Lambda(\alpha^2+1)^{2}b^{2\gamma(\frac{n-1}{n-3})}}{(n-1)(\alpha^2-n)}\rho^{2(1-\gamma(\frac{n-1}{n-3}))}
+\frac{mb^{(\frac{4\gamma}{n-3})}}{\rho^{n-2}}\rho^{\gamma(n-1-\frac{4}{n-3})}
\nonumber \\
&&-\frac{2q^{2}(\alpha^2+1)^{2}b^{-2\gamma\left(n-2-\frac{2}{n-3}\right)}}{(n-1)(\alpha^2+n-2)\rho^{2[(n-2)(1-\gamma)+
\frac{2\gamma}{n-3}]}} ,
\label{V1}\\
H(\rho)&=&\left(\frac{b}{\rho}\right)^{\frac{(n-5)\gamma}{n-3}},
\label{H1}\\
\Phi(\rho)&=&\left(\frac{b}{\rho}\right)^{\frac{2(n-1)\gamma}{n-3}}.
\label{Phi1}
\end{eqnarray}
The electromagnetic field becomes
\begin{eqnarray}
F_{\phi\rho}&=&\frac{q
lb^{(3-n)\gamma}}{\rho^{(n-3)(1-\gamma)+2}}.\label{Ftr2}
\end{eqnarray}
It is worthwhile to note that the scalar field $\Phi(\rho)$ and
the electromagnetic field $F_{\phi\rho}$ become zero as
$\rho\rightarrow\infty$. As one can see from Eqs. (\ref{U1}) and
(\ref{V1}), the solutions are ill-defined for $\alpha=\sqrt{n}$
with $\Lambda\neq0$ (corresponding to $\omega=-3(n+3)/4n$). In the
limiting case $\alpha\rightarrow 0$ ($\omega\rightarrow \infty$),
our solutions restore those presented in \cite{Deh2} for magnetic
branes in Einstein-Maxwell theory.

Next, we study the properties of the solutions. First of all, we
seek for curvature singularities. One can easily check that the
Kretschmann scalar, $R_{\mu \nu \lambda \kappa }R^{\mu \nu \lambda
\kappa }$, diverges at $\rho =0$ and therefore one might think
that there is a curvature singularity located at $\rho =0$.
However, a profound look at the metric reveals that the spacetime
will never achieve $\rho =0$. The function $V(\rho )$ is negative
for $\rho <r_{+}$ and positive for $\rho
>r_{+}$, where $r_{+}$ is the largest root of $V(\rho )=0$. Indeed, $g_{\rho \rho }$ and $g_{\phi \phi }$ are related by $V(\rho
)=g_{\rho \rho }^{-1}\propto l^{-2}g_{\phi \phi }$, and therefore
when $g_{\rho \rho }$ becomes negative (which occurs for $\rho
<r_{+}$) so does $g_{\phi
\phi }$. This leads to apparent change of signature of the metric from $%
(n-1)+$ to $(n-2)+$ as one extends the spacetime to $\rho <r_{+}$.
This indicates that we are using an incorrect extension. To get
rid of this incorrect extension, we introduce the new radial
coordinate $r$ as
\[
r^{2}=\rho ^{2}-r_{+}^{2}\Rightarrow d\rho ^{2}=\frac{r^{2}}{r^{2}+r_{+}^{2}}%
dr^{2}.
\]
With this new coordinate, the metric (\ref{met2}) becomes
\begin{eqnarray}
ds^{2}
&=&-\frac{r^{2}+r_{+}^{2}}{l^{2}}H^{2}(r)dt^{2}+l^{2}U(r)d\phi
^{2}
\nonumber \\
&&+\frac{r^{2}dr^{2}}{(r^{2}+r_{+}^{2})V(r)}+\frac{r^{2}+r_{+}^{2}}{l^{2}}%
H^{2}(r)dX^{2},  \label{met3}
\end{eqnarray}
where the coordinates $r$ assumes the values $0\leq r <\infty$,
and $U(r)$, $V(r)$, $H(r)$ and $\Phi(r)$ are now given as
\begin{eqnarray}
U(r)&=&\frac{2\Lambda(\alpha^2+1)^{2}b^{2\gamma(\frac{n-5}{n-3})}}{(n-1)(\alpha^2-n)}(r^2+r_{+}^2)^{(1-\frac{\gamma(n-5)}{n-3})}+
\frac{mb^{(\frac{-4\gamma}{n-3})}}{(r^2+r_{+}^2)^{(n-2)/2}}(r^2+r_{+}^2)^{\left[n-1+\frac{4}{n-3}\right]\gamma/2}\nonumber \\
&&-\frac{2q^{2}(\alpha^2+1)^{2}b^{-2\gamma\left(n-2+\frac{2}{n-3}\right)}}{(n-1)\left(\alpha^2+n-2\right)(r^2+r_{+}^2)^{[(n-2)(1-\gamma)-
\frac{2\gamma}{n-3}]}},
 \label{Ur1}\\
V(r)&=&\frac{2\Lambda(\alpha^2+1)^{2}b^{2\gamma(\frac{n-1}{n-3})}}{(n-1)(\alpha^2-n)}(r^2+r_{+}^2)^{[1-\gamma(\frac{n-1}{n-3})]}
+\frac{mb^{(\frac{4\gamma}{n-3})}}{(r^2+r_{+}^2)^{(n-2)/2}}(r^2+r_{+}^2)^{\left[n-1-\frac{4}{n-3}\right]\gamma/2}
\nonumber \\
&&-\frac{2q^{2}(\alpha^2+1)^{2}b^{-2\gamma\left(n-2-\frac{2}{n-3}\right)}}{(n-1)(\alpha^2+n-2)(r^2+r_{+}^2)^{[(n-2)(1-\gamma)+
\frac{2\gamma}{n-3}]}} ,
\label{Vr1}\\
H(r)&=&\left(\frac{b}{\sqrt{r^2+r_{+}^2}}\right)^{\frac{(n-5)\gamma}{n-3}},
\label{Hr1}\\
\Phi(r)&=&\left(\frac{b}{\sqrt{r^2+r_{+}^2}}\right)^{\frac{2(n-1)\gamma}{n-3}}.
\label{Phir1}
\end{eqnarray}
The asymptotic behavior of the spacetime (\ref{met3}) is neither
flat nor (anti)-de Sitter. It is easy to check that the
Kretschmann scalar does not diverge in the range $0\leq r<\infty
$. However, the spacetime has a conic geometry and has a conical
singularity at $r=0$. Indeed, there is a conical singularity at
$r=0$ since:
\begin{eqnarray}\label{metr0a}
\lim_{r\rightarrow 0}\frac{1}{r}\sqrt{\frac{g_{\phi \phi
}}{g_{rr}}}\neq 1.
\end{eqnarray}
That is, as the radius $r$ tends to zero, the limit of the ratio ``\textrm{%
circumference/radius}'' is not $2\pi $ and therefore the spacetime
has a conical singularity at $r=0$. The canonical singularity can
be removed if one identifies the coordinate $\phi$ with the period
\begin{equation}
\textrm{Period}_{\phi}=2 \pi \left(\lim_{r\rightarrow
0}\frac{1}{r}\sqrt{\frac{g_{\phi \phi }}{g_{rr}}} \right)^{-1}=2
\pi (1-4 \mu), \label{period}
\end{equation}
where $\mu$ is given by
\begin{equation}
\mu=\frac{1}{4}\left[1-\left(\frac{1}{2}\frac{ml(\alpha ^{2}+n-2)}{\alpha ^{2}+1}r_{+}^{(n-1)(\gamma -1)}+%
\frac{2(1+\alpha ^{2})}{(\alpha ^{2}-n)}\Lambda lb^{2\gamma
}r_{+}^{1-2\gamma }\right)^{-1}\right]. \label{mu}
\end{equation}
By the above analysis, we conclude that near the origin $r=0$, the
metric (\ref{met3}) describes a spacetime which has a conical
singularity at $r=0$ with a deficit angle $\delta=8 \pi \mu$,
which is proportional to the brane tension at $r=0$
\cite{ranjbar}. In order to investigate the effects of the scalar
field on the the deficit angle, we plot in Figs. 1 and 2 the
deficit angle $\delta$ versus $\alpha$. These figures show that
for  $\Lambda<0$ the deficit angle $\delta$ of the spacetime
increases with increasing $\alpha$ while for $\Lambda>0$ the
deficit angle decreases with increasing $\alpha$. Of course, one
may ask for the completeness of the spacetime with $r\geq 0$ (or
$\rho \geq r_{+}$). It is easy to see that the spacetime described
by Eq. (\ref{met3}) is both null and timelike geodesically
complete as in the case of four-dimensional solutions
\cite{Lem2,Hor3}. In fact, one can show that every null or
timelike geodesic starting from an arbitrary point can either
extend to infinite values of the affine parameter along the
geodesic or end on a singularity at $r=0$ \cite{SDR}.

Now we investigate the casual structure of the spacetime. As one
can see from Eqs. (\ref{Ur1})-(\ref{Vr1}), there is no solution
for $\alpha =\sqrt{n}$ with a quadratic
potential for the scalar field ($\Lambda \neq 0$%
). The cases with $\alpha >\sqrt{n}$ and $\alpha <\sqrt{n}$ should
be considered separately. For $\alpha >\sqrt{n}$, as $r$ goes to
infinity, the dominant term in Eqs. (\ref{Ur1})-(\ref{Vr1}) is the
second term, and therefore the functions $U(r)$ and $V(r)$ are
positive in the whole spacetime, despite the sign of the
cosmological constant $\Lambda $, and is zero at $r=0$. Thus, the
solution given by Eqs. (\ref{met3})-(\ref{Vr1}) exhibits a
spacetime with conic singularity at $r=0 $. For $\alpha
<\sqrt{n}$, the dominant term for large values of $r$\ is the
first term, and therefore the functions $U(r)$ and $V(r)$ are
positive in the whole spacetime for $\Lambda<0$. In this case the
solution represents a spacetime with conic singularity at $r=0$.
The solution is not acceptable for $\alpha <\sqrt{n}$ with
$\Lambda>0$, since the functions $U(r)$ and $V(r)$ are negative
for large values of $r$.
\begin{figure}[tbp]
\epsfxsize=7cm \centerline{\epsffile{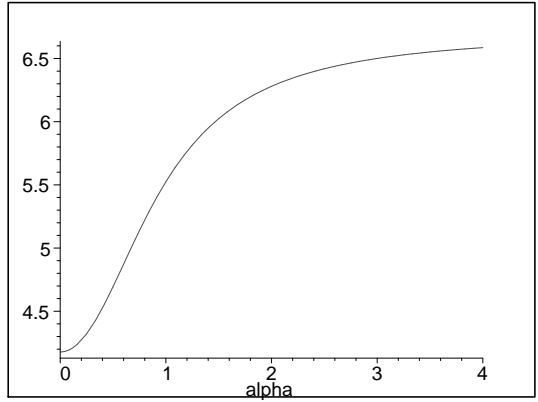}} \caption{Deficit
angle $\delta$ versus $\alpha$  for $n=4$, $b=1$, $l=1$,
$\Lambda=-6$, $m=0.3$ and $r_{+}=0.8$.} \label{figure1}
\end{figure}
\begin{figure}[tbp]
\epsfxsize=7cm \centerline{\epsffile{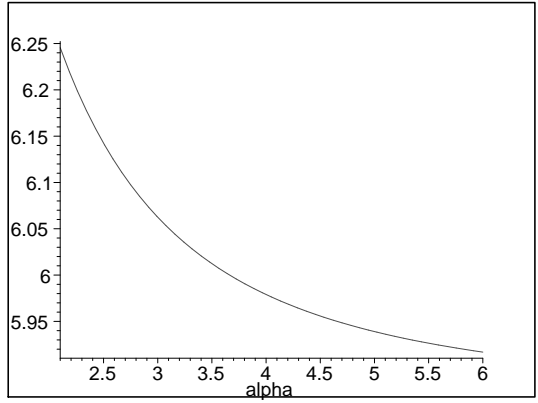}} \caption{Deficit
angle $\delta$ versus $\alpha$  for $n=4$, $b=1$, $l=1$,
$\Lambda=+6$, $m=0.3$ and $r_{+}=0.8$.} \label{figure2}
\end{figure}
\section{Spinning magnetic Branes in BD theory\label{Spin}}
In this section, we would like to endow our spacetime metric
(\ref{met2}) with a rotation. First, we consider the solutions
with one rotation parameter. In order to add an angular momentum
to the spacetime, we perform the following rotation boost in the
$t-\phi $ plane
\begin{equation}
t\mapsto \Xi t-a\phi ,\hspace{0.5cm}\phi \mapsto \Xi \phi -\frac{a}{%
l^{2}}t,  \label{Tr}
\end{equation}
where $a$ is a rotation parameter and $\Xi =\sqrt{1+a^{2}/l^{2}}$.
Substituting Eq. (\ref{Tr}) into Eq. (\ref{met2}) we obtain
\begin{eqnarray}
d\bar{s}^{2}&=&-\frac{r^2+r_{+}^2}{l^{2}}H^{2}(r )\left( \Xi
dt-ad\phi \right) ^{2}+\frac{r^{2}dr^{2}}{(r^{2}+r_{+}^{2})V(r)}\nonumber \\
&&+l^{2}U(r )\left( \frac{a}{l^{2}}dt-\Xi d\phi \right)
^{2}+\frac{r^2+r_{+}^2}{l^{2}}H^{2}(r )dX^{2}{,}
 \label{met4}
\end{eqnarray}
where $U(r)$, $V(r)$ and $H(r)$ are given in Eqs.
(\ref{Ur1})-(\ref{Hr1}). The non-vanishing electromagnetic field
components become
\begin{eqnarray}
F_{\phi r}&=&\frac{ql \Xi
b^{(3-n)\gamma}}{(r^2+r_{+}^2)^{[(n-3)(1-\gamma)+2]/2}},\hspace{0.7cm}
F_{tr }=-\frac{a}{\Xi l^{2}}F_{\phi r}.
\end{eqnarray}
The transformation (\ref{Tr}) generates a new metric, because it
is not a permitted global coordinate transformation. This
transformation can be done locally but not globally \cite{Sta}.
Therefore, the metrics (\ref{met3}) and (\ref{met4}) can be
locally mapped into each other but not globally, and so they are
distinct. Note that this spacetime has no horizon and curvature
singularity. However, it has a conical singularity at $r=0$.

Second, we study the rotating solutions with a complete set of
rotation parameters. The rotation group in $n+1$ dimensions is
$SO(n)$ and therefore the number of independent rotation
parameters is $[n/2]$, where $[x]$ is the integer part of $x$. We
now generalize the above metric given in Eq. (\ref{met4}) with
$k\leq \lbrack n/2]$ rotation parameters. This generalized
solution can be written as
\begin{eqnarray}
ds^{2} &=&-\frac{r^2+r_{+}^2}{l^{2}}H^{2}(r)\left( \Xi dt-{{%
\sum_{i=1}^{k}}}a_{i}d\phi ^{i}\right) ^{2}+U(r)\left( \sqrt{\Xi ^{2}-1}dt-%
\frac{\Xi }{\sqrt{\Xi ^{2}-1}}{{\sum_{i=1}^{k}}}a_{i}d\phi
^{i}\right) ^{2}
\nonumber \\
&&+\frac{r^{2}dr^{2}}{(r^{2}+r_{+}^{2})V(r)}+\frac{r^2+r_{+}^2}{l^{2}(\Xi
^{2}-1)}H^{2}(r){\sum_{i<j}^{k}}(a_{i}d\phi _{j}-a_{j}d\phi
_{i})^{2}+\frac{r^2+r_{+}^2}{l^{2}}H^{2}(r)dX^{2}, \label{met5}
\end{eqnarray}
where $\Xi =\sqrt{1+\sum_{i}^{k}a_{i}^{2}/l^{2}}$, $dX^{2}$ is the
Euclidean metric on the $(n-k-1)$-dimensional submanifold with
volume $V_{n-k-1}$. The functions $U(r)$, $V(r)$ and $H(r)$ are
those presented in Eqs. (\ref{Ur1})-(\ref{Hr1}). The non-vanishing
components of electromagnetic field tensor are
\begin{eqnarray}
F_{tr }=-\frac{\Xi^2-1}{\Xi a_{i} }F_{\phi^{i}
r}=-\frac{q\sqrt{\Xi^2-1}b^{(3-n)\gamma}}{(r^2+r_{+}^2)^{[(n-3)(1-\gamma)+2]/2}}.
\end{eqnarray}
The corresponding gauge potential of the general solution
(\ref{met5}) is given by
\begin{equation}
A_{\mu }=\frac{qb^{(3-n)\gamma }}{\Gamma (r^2+r_{+}^2)^{\Gamma/2
}}
\left( \sqrt{\Xi ^{2}-1}\delta _{\mu }^{t}-\frac{\Xi }{\sqrt{\Xi ^{2}-1}}%
a_{i}\delta _{\mu }^{i}\right) \hspace{0.5cm}{\text{(no sum on
}i\text{)}}. \label{A3}
\end{equation}
where $\Gamma =(n-3)(1-\gamma )+1$. Again this spacetime has no
horizon and curvature singularity. However, it has a conical
singularity at $r=0$. One should note that these solutions reduce
to those discussed in \cite{Deh2}, in the absence of scalar field
($\alpha =\gamma =0$) and those presented in \cite{Deh4} for
$n=3$.

\section{Counterterm method and Conserved Quantities \label{Cons}}
The action (\ref{act1}) does not have a well-defined variational
principle, since one encounters a total derivative that produces a
surface integral involving the derivative of $\delta g_{\mu\nu}$
normal to the boundary. These normal derivative terms do not
vanish by themselves, but are cancelled by the variation of the
boundary term
\begin{equation}
I_{b}=- \frac{1}{8\pi}\int_{\partial \mathcal{M}}{d^{n}x\sqrt{h} K
\Phi},\label{actb}
\end{equation}
where $h$ and $K$ are the determinant of the induced metric and
the trace of extrinsic curvature of boundary. In general the
action $I_{G}+I_b$ , is divergent when evaluated on the solutions.
A systematic method of dealing with this divergence for
asymptotically AdS solutions of Einstein gravity is through the
use of the counterterms method inspired by the anti-de Sitter
conformal field theory  correspondence (AdS/CFT) \cite{Mal}.
However, in the presence of a non-trivial BD scalar field with
potential $V(\Phi)
=2\Lambda \Phi^2$, the spacetime may not behave as either dS ($\Lambda >0$) or AdS ($%
\Lambda <0$). In fact, it has been shown that with the exception
of a pure cosmological constant potential, where $\alpha =0$, no
de Sitter or anti-de Sitter static spherically symmetric solution
exist for one Liouville-type dilaton potential \cite{PW}. But, as
in the case of asymptotically AdS spacetimes, according to the
domain-wall/QFT (quantum field theory) correspondence \cite{Sken},
there may be a suitable counterterm for the stress energy tensor
which removes the divergences. In this paper, we deal with the
spacetimes with zero curvature boundary, and therefore all the
counterterm containing the curvature invariants of the boundary
are zero. Thus, the suitable counterterm action may be written
\begin{equation}
I_{ct}=- \frac{1}{8\pi}\int_{\partial \mathcal{M}}{d^{n}x\sqrt{h}
\frac{(n-1)}{l_{\mathrm{eff}}}},\label{actct}
\end{equation}
where $l_{\mathrm{eff}}$ is given by
\begin{equation}
l_{\mathrm{eff}}^{2}=\frac{(n-1)(\alpha ^{2}-n)}{2\Lambda \Phi^3}.
\label{leff}
\end{equation}
In the limiting case $\alpha \rightarrow0$$(\Phi=1)$, the
effective $l_{\mathrm{eff}}^{2}$ of Eq. (\ref{leff}) reduces to
$l^{2}=-n(n-1)/2\Lambda $ of the AdS spacetimes. Using
(\ref{actb}) and (\ref{actct}) the finite stress-energy tensor in
$(n+1)$-dimensional BDM theory may be written as
\begin{equation}
T^{ab}=\frac{1}{8\pi }\left[\left( K^{ab}-K h^{ab}\right)\Phi+\frac{n-1}{l_{%
\mathrm{eff}}}h ^{ab}\right] ,  \label{Stres}
\end{equation}
The first two terms in Eq. (\ref{Stres}) are the variation of the
action (\ref {actb}) with respect to $h _{ab}$, and the last term
is the counterterm which removes the divergences. Note that the
counterterm has the same form as in the case of asymptotically AdS
solutions with zero curvature boundary, where $l$ is replaced by
$l_{\mathrm{eff}}$. To compute the
conserved charges of the spacetime, one should choose a spacelike surface $%
\mathcal{B}$ in $\partial \mathcal{M}$ with metric $\sigma _{ij}$,
and write the boundary metric in ADM (Arnowitt-Deser-Misner) form:
\[
\gamma _{ab}dx^{a}dx^{a}=-N^{2}dt^{2}+\sigma _{ij}\left( d\varphi
^{i}+V^{i}dt\right) \left( d\varphi ^{j}+V^{j}dt\right) ,
\]
where the coordinates $\varphi ^{i}$ are the angular variables
parameterizing the hypersurface of constant $r$ around the origin,
and $N$ and $V^{i}$ are the lapse and shift functions,
respectively. When there is a Killing vector field $\mathcal{\xi
}$ on the boundary, then the quasilocal conserved quantities
associated with the stress tensors of Eq. (\ref{Stres}) can be
written as
\begin{equation}
Q(\mathcal{\xi )}=\int_{\mathcal{B}}d^{n-1} x \sqrt{\sigma }T_{ab}n^{a}%
\mathcal{\xi }^{b},  \label{charge}
\end{equation}
where $\sigma $ is the determinant of the metric $\sigma _{ij}$, $\mathcal{%
\xi } $ and $n^{a}$ are the Killing vector field and the unit
normal vector on the boundary $\mathcal{B}$. For boundaries with
timelike ($\xi =\partial /\partial t $) and rotational
($\varsigma_{i} =\partial /\partial \phi^{i} $) Killing vector
fields one obtains the quasilocal mass and components of total
angular momenta as
\begin{eqnarray}
M &=&\int_{\mathcal{B}}d^{n-1} x \sqrt{\sigma }T_{ab}n^{a}\xi
^{b},
\label{Mastot} \\
J_{i} &=&\int_{\mathcal{B}}d^{n-1} x \sqrt{\sigma
}T_{ab}n^{a}\varsigma_{i} ^{b},  \label{Angtot}
\end{eqnarray}
provided the surface $\mathcal{B}$ contains the orbits of
$\varsigma $. These quantities are, respectively, the conserved
mass and angular momenta of the system enclosed by the boundary
$\mathcal{B}$. Note that they will both be dependent on the
location of the boundary $\mathcal{B}$ in the spacetime, although
each is independent of the particular choice of foliation
$\mathcal{B}$ within the surface $\partial \mathcal{M}$. Now we
are in a position to calculate conserved quantities of the
solutions. Denoting the volume of the hypersurface boundary at constant $t$ and $r$ by $%
V_{n-1}=(2\pi )^{k}\Sigma _{n-k-1}$, the mass and angular momentum
per unit volume $V_{n-1}$ of the branes ($\alpha <\sqrt{n}$) can
be calculated through the use of Eqs. (\ref{Mastot}) and
(\ref{Angtot}). We find
\begin{eqnarray}
{M} &=&\frac{b^{(n-1)\gamma }}{16\pi l^{n-2}}\left(
\frac{(n-\alpha ^{2})\Xi
^{2}-(n-1)}{1+\alpha ^{2}}\right) m,  \label{M1} \\
J_{i} &=&\frac{b^{(n-1)\gamma }}{16\pi l^{n-2}}\left( \frac{n-\alpha ^{2}}{%
1+\alpha ^{2}}\right) \Xi ma_{i}.  \label{J1}
\end{eqnarray}
For $a_{i}=0$ ($\Xi =1$), the angular momentum per unit volume
vanishes, and therefore $a_{i}$'s are the rotational parameters of
the spacetime. Comparing the conserved quantities calculated in
this section with those obtained in \cite{SDR}, we find that they
are invariant under the conformal transformation (\ref{conform2}).
In the particular case $n=3$, these conserved quantities reduce to
the conserved quantities of the magnetic rotating black string
obtained in Ref. \cite{Deh4}, and in the absence of dilaton field
($\alpha =0=\gamma$) they reduce to those presented in Ref.
\cite{Deh2}.

Finally, we calculate the electric charge of the solutions. To
determine the electric field we should consider the projections of
the electromagnetic field tensor on special hypersurfaces. The
normal to such hypersurfaces for the spacetimes with a
longitudinal magnetic field is
\[
u^{0}=\frac{1}{N},\text{ \ }u^{r}=0,\text{ \
}u^{i}=-\frac{V^{i}}{N},
\]
and the electric field is $E^{\mu }=g^{\mu \rho }F_{\rho \nu
}u^{\nu }$. Then the electric charge per unit volume $V_{n-1}$ can
be found by calculating the flux of the electric field at
infinity, yielding
\begin{equation}
{Q}=\frac{\sqrt{\Xi ^{2}-1}q}{4\pi l^{n-2}}.
\end{equation}
It is worth noting that the electric charge of the system per unit
volume is proportional to the magnitude of rotation parameters and
is zero for the case of a static solution. This result is expected
since now, besides the magnetic field along the $\phi ^{i}$
coordinates, there is also a radial electric field ($F_{t\rho}\neq
0$). To give a physical interpretation for the appearance of the
net electric charge, we first consider the static spacetime. The
magnetic field source can be interpreted as composed of equal
positive and negative charge densities, where one of the charge
density is at rest and the other one is spinning. Clearly, this
system produce no electric field since the net electric charge
density is zero, and the magnetic field is produced by the
rotating electric charge density. Now, we consider the spining
solutions. From the point of view of an observer at rest relative
to the source ($S$), the two charge densities are equal, while
from the point of view of an observe $S^{\prime }$ that follows
the intrinsic rotation of the spacetime, the positive and negative
charge densities are not equal, and therefore the net electric
charge of the spacetime is not zero.
\section{Conclusion \label{Con}}
To conclude, we found a new class of magnetic solutions in
Brans-Dicke-Maxwell theory in the presence of a quadratic
potential for the scalar field. These solutions are neither
asymptotically flat nor (anti)-de Sitter. This class of solutions
represents an $(n+1)$-dimensional spacetime with a longitudinal
magnetic field generated by a static magnetic brane. We
investigated the properties of these solutions and found that
these solutions have no curvature singularity and no horizons, but
have conic singularity at $r=0$ with a deficit angle $\delta$.
Then, we generalized our solutions to the case of spining magnetic
brane with $k\leq \lbrack n/2]$ rotation parameters. For the
spinning brane, when the rotation parameters are nonzero, the
brane has a net electric charge density which is proportional to
the magnitude of the rotation parameters given by $\sqrt{\Xi
^{2}-1}$. We obtained the the conserved quantities of these
spacetimes  through the use of counterterm method inspired by
(A)dS/CFT correspondence.

\acknowledgments{This work has been supported financially by
Research Institute for Astronomy and Astrophysics of Maragha,
Iran.}


\begin{thebibliography}{99}
\bibitem{Rie} A. G. Riess, et al., Astron. J.  {\textbf{116}}, 1009 (1998);\\
  S. Perlmutter, et al.,   Astrophys. J.  {\bf517}, 565 (1999);\\
  S. Perlmutter, et al.,   Astrophys. J.  {\bf598}, 102 (2003); \\
 P. de Bernardis, et al.,   Nature  {\bf404}, 955 (2000).
\bibitem{BD} C. Brans and R. H. Dicke, Phys. Rev. $\bold{124}$,  925 (1961).

\bibitem{Will} C. M. Will, \textit{Theory and Experiment in Gravitational
Physics}, (Cambridge University Press, Cambridge, 1993).

\bibitem{Bert} B. Bertotti, L. Iess and P. Tortora, Nature, 425 (2003)
374.

\bibitem{Aca} V. Acquaviva, L. Verde, JCAP 12 (2007) 001.

\bibitem{Sen} A. Sen, Phys. Rev. Lett. $\bold{69}$, 1006 (1992);\\
 G. W. Gibbons and K. Maeda, Ann. Phys. (N.Y.) $\bold{167}$, 201 (1986);\\
 V. Frolov, A. Zelinkov and U. Bleyer, Ann. Phys. (Leipzig) $\bold{44}$, 371 (1987).

\bibitem{Brans} C. H. Brans, Phys. Rev. $\bold{125}$, 2194 (1962).
\bibitem{Hawking} S. W. Hawking, Commun. Math. Phys. $\bold{25}$, 167 (1972).
\bibitem{Cai1} R. G. Cai and Y. S. Myung, Phys. Rev. D. $\bold{56}$, 3466 (1997).

\bibitem{Cai2} R. G. Cai, J.Y. Ji and K. S. Soh, Phys. Rev. D {\bf57}, 6547 (1998).

\bibitem{Kim} H. Kim, Phys. Rev. D, $\bold{60}$, 024001 (1999).

\bibitem{Deh1} M. H. Dehghani, J. Pakravan, and S. H. Hendi, Phys. Rev. D {\bf74}, 104014 (2006); \\
S. H. Hendi, J. Math. Phys. {\bf49}, 082501 (2008).

\bibitem{Gao} H. Kim, Nuovo Cim. B {\bf112}, 329 (1997).

\bibitem {SheyAla} A. Sheykhi, H. Alavirad,  IJMPD
Vol. 18, No. 9 (2009) in press; \\
 A. Sheykhi, M. M. Yazdanpanah, Phys. Lett. B 679 (2009) 311.

\bibitem {TV} T. Vachaspati, A. Vilenkin, Phys. Rev. Lett. 67 (1991) 1057.
\bibitem {AV} A. Vilenkin, E.P.S. Shellard, \textit{Cosmic Strings and Other
Topological Defects}, Cambridge University Press, New York, 1994.


\bibitem{Vil}  A. Vilenkin, Phys. Rev. D {\bf 23}, 852 (1981);\\
 W. A. Hiscock, Phys. Rev. D. {\bf 31}, 3288 (1985); \\
 D. Harari and P. Sikivie, Phys. Rev. D {\bf 37},  3438 (1988);\\
 A. D. Cohen and D. B. Kaplan, Phys. Lett. B {\bf 215}, 65 (1988); \\
 R. Gregory, Phys. Rev. D. {\bf 215}, 663 (1988).

\bibitem{Ban} A. Banerjee, N. Banerjee, and A. A. Sen, Phys. Rev. D {\bf 53},  5508 (1996); \\
 M. H. Dehghani and T. Jalali, Phys. Rev. D {\bf 66}, 124014 (2002) ;\\
 M. H. Dehghani and A. Khodam-Mohammadi, Can. J. Phys. {\bf 83}, 229 (2005).

\bibitem{Bon} W. B. Bonnor, Proc. Roy. S. London A {\bf67}, 225 (1954); \\
A. Melvin, Phys. Lett. {\bf8}, 65 (1964).

\bibitem{Lem1}  O. J. C. Dias and J. P. S. Lemos, J. High Energy Phys. {\bf01}, 006 (2002).

\bibitem{Lem2}  O. J. C. Dias and J. P. S. Lemos, Class. Quant. Gravit. {\bf19}, 2265 (2002).

\bibitem{Deh2}  M. H. Dehghani, Phys. Rev. D {\bf 69}, 044024 (2004).
\bibitem{Deh3}  M. H. Dehghani, Phys. Rev. D {\bf 69}, 064024
(2004);\\ M. H. Dehghani,  N. Bostani and S. H. Hendi, Phys. Rev.
D {\bf78}, 064031 (2008).

\bibitem{Wit2}  E. Witten, Nucl. Phys. B {\bf 249},  557 (1985);\\
 P. Peter, Phys. Rev. D {\bf 49},  5052 (1994).

\bibitem{Moss}  I. Moss and S. Poletti, Phys. Lett. B {\bf199}, 34 (1987).

\bibitem{Fer}  C. N. Ferreira, M. E. X. Guimaraes and J. A. Helayel-Neto,
Nucl. Phys. B {\bf581}, 165 (2000).



\bibitem{Deh4}  M. H. Dehghani, Phys. Rev. D {\bf 71},  064010 (2005).

\bibitem{SDR}  A. Sheykhi, M. H. Dehghani, N. Riazi, Phys. Rev. D {\bf75}, 044020 (2007) .
\bibitem{DSH} M. H. Dehghani, A. Sheykhi and S. H. Hendi, Phys. Lett. B
{\bf659},  476 (2008).

\bibitem{Sheykhi}  A. Sheykhi, Phys. Lett. B {\bf672}, 101 (2009).



\bibitem{Dia}  O. J. C. Dias and J. P. S. Lemos, Phys. Rev. D {\bf66},  024034 (2002).

\bibitem{Sen1}  A. A. Sen, Phys. Rev. D {\bf 60}, 067501 (1999).

\bibitem{ranjbar} S. Randjbar-Daemi and V. Rubakov, J. High Energy Phys. {\bf10}, 054 (2004);\\
 H. M. Lee and A. Papazoglou, Nucl. Phys. B {\bf705}, 152
(2005).
\bibitem{Hor3}  J. H. Horne and G. T. Horowitz, Nucl. Phys. B {\bf368}, 444
(1992).

\bibitem{Sta} J. Stachel, Phys. Rev. D {\bf26}, 1281 (1982).

\bibitem{Mal}  J. Maldacena, Adv. Theor. Math. Phys., {\bf 2}, 231
(1998);\\
E. Witten, Adv. Theor. Math. Phys. {\bf 2}, 253 (1998); \\
O. Aharony, S. S. Gubser, J. Maldacena, H. Ooguri and Y. Oz, Phys.
Rep. {\bf 323}, 183 (2000);\\ V. Balasubramanian and P. Kraus,
Commun. Math. Phys. {\bf 208,} 413 (1999).

\bibitem{PW}  S. J. Poletti and D. L. Wiltshire, Phys. Rev. D {\bf 50}, 7260
(1994).

\bibitem{Sken}  H. J. Boonstra, K. Skenderis, and P. K. Townsend, J. High
Energy Phys. {\bf 01}, 003 (1999);\\
 K. Behrndt, E. Bergshoeff, R. Hallbersma and J. P. Van der Scharr, Class. Quant. Gravit. {\bf
16}, 3517 (1999);\\
 R. G. Cai and N. Ohta, Phys. Rev. D {\bf 62}, 024006 (2000).

\end{thebibliography}
\end{document}